
\magnification=1200
\def\tr#1{{\rm tr #1}}

\def\f#1#2{{\textstyle{#1\over #2}}}

\def\next{\hfil\break\noindent}

\font\title=cmbx12

{\title
\centerline{On the nature of singularities}
\centerline{in plane symmetric scalar field cosmologies}}

\vskip 2cm

\noindent
Alan D. Rendall\footnote{${}^1$}{
Max-Planck-Institut f\"ur Astrophysik,
Karl-Schwarzschildstr. 1,
Postfach 1523,
85740 Garching,
Germany.}

\vskip 2cm

\noindent
{\bf Abstract}

\vskip 10pt\noindent
The nature of the initial singularity in spatially
compact plane symmetric scalar field cosmologies is investigated.
It is shown that this singularity is crushing and velocity
dominated and that the Kretschmann scalar diverges uniformly
as it is approached. The last fact means in particular that a
maximal globally hyperbolic spacetime in this class cannot be
extended towards the past through a Cauchy horizon. A subclass
of these spacetimes is identified for which the singularity is
isotropic.

\vskip 2cm

\noindent
{\bf 1. INTRODUCTION}

The nature of singularities in general solutions of the Einstein
equations is a subject about which much remains to be learned.
Various classes of singularities have been defined which represent
possible models for general behaviour. Examples are curvature
singularities, crushing singularities[1], velocity dominated
singularities[2] and isotropic singularities[3]. In this paper
spacetimes belonging to one of the simplest classes of
inhomogeneous cosmologies will be examined in order to get
as much information as possible about their singularities
and to test the applicability of the models just mentioned.

The spacetimes considered in the following are solutions of
the Einstein equations coupled to a massless scalar field in
the standard way. Thus, if $\phi$ denotes the scalar field
they are solutions of
$$G_{\alpha\beta}=8\pi[\nabla_\alpha\phi\nabla_\beta\phi
-\f12(\nabla^\gamma\phi\nabla_\gamma\phi)g_{\alpha\beta}].\eqno(1)$$
The Bianchi identities imply that $\phi$ satisfies the wave equation.
These spacetimes are further assumed to be plane symmetric.
Plane symmetric solutions of the Einstein equations with a scalar
field as matter source have been discussed by Tabensky and Taub [4].
In fact their paper is on stiff fluids but, as they show, it is
possible to transform between these two matter models under rather
general circumstances. They write the field equations in a
particularly simple form. If the gradient of the area of the
orbits is everywhere timelike then these equations
can be simplified further. This condition will be assumed in the
following. It has been shown elsewhere that for appropriate
boundary conditions it is automatically fulfilled unless the
spacetime is flat[5]. Tabensky and Taub show that the only
non-trivial equation to be solved is the linear hyperbolic equation
$$\phi_{tt}+t^{-1}\phi_t=\phi_{xx}\eqno(2)$$
When this has been done a quantity $\Omega$ is obtained by
integrating the ordinary differential equation
$$\Omega_t=t(\phi_t^2+\phi_x^2)\eqno(3)$$
This can be done starting on an initial hypersurface of constant
$t$. In order that all Einstein equations should be satified
the constraint equation
$$\Omega_x=2t\phi_t\phi_x\eqno(4)$$
should hold on the initial hypersurface. The spacetime metric is
$$ds^2=t^{-1/2}e^\Omega(-dt^2+dx^2)+t(dy^2+dz^2)\eqno(5)$$
Here $t$ belongs to the interval $(0,\infty)$.
To avoid spurious singularities it is assumed that the
spacetime is spatially compact. This can be arranged by demanding
that the coordinates $x$, $y$ and $z$ be periodic. The periodicity
of $y$ and $z$ plays no significant role in the following but
the periodicity of $x$ means that $\phi$ and $\Omega$ (which only
depend on $t$ and $x$) are required to be periodic in $x$.

The initial value problem for data given on the hypersurface
$t=t_0>0$ can be solved as follows. An initial data set consists of
periodic functions $\phi_0$, $\phi_1$ and $\Omega_0$ which
satisfy the equation
$$(\Omega_0)_x=2t_0\phi_1(\phi_0)_x.\eqno(6)$$
For simplicity they will be assumed to be $C^\infty$ although
the arguments which follow can also be carried through when
these functions have an appropriate finite degree of
differentiability. A solution is sought with $\phi(t_0,x)=\phi_0(x)$,
$\phi_t(t_0,x)=\phi_1(x)$ and $\Omega(t_0,x)=\Omega_0(x)$.
Under these conditions (6) is just the constraint equation
(4) on the hypersurface $t=t_0$. To construct the solution
first solve the linear hyperbolic equation (2) on the time
interval $(0,\infty)$ with initial data $\phi_0$ and $\phi_1$.
Standard theory ensures the existence of a unique $C^\infty$
solution $\phi$. Then  $\Omega$ may be determined by integrating
(3) with initial value $\Omega_0$ for each fixed value of $x$.

The solutions of the initial value problem have an apparent
singularity at $t=0$. The aim of the following is to show that
this is a true singularity (i.e. that the spacetime cannot be
extended through it) and to obtain more detailed information
about its nature. In Section 2 it is shown that $t=0$ is
always a curvature singularity and that the Kretschmann
scalar $R^{\alpha\beta\gamma\delta}R_{\alpha\beta\gamma\delta}$
blows up uniformly as $t\to 0$. The consequences for strong
cosmic censorship are discussed. In Section 3 the singularity
is shown to be crushing and it is concluded that a neighbourhood
of the singularity can be foliated by constant mean curvature
hypersurfaces. An asymptotic expansion for the solution in a
neighbourhood of $t=0$ is obtained in Section 4 which shows
in particular that the singularity is velocity dominated.
In the final section a sufficient condition is given for the
singularity to be isotropic.

Note that the simplification of the field equations which
allows the analysis which follows to be carried out depends
very much on the plane symmetry and the fact that the matter
content of spacetime is described by a massless scalar field.
If plane symmetry is replaced by spherical symmetry or if
the massless scalar field is replaced by almost any other
kind of matter, then the equations for the matter fields,
the equation for $\Omega$ and the equation for the area
of the orbits are all coupled. The property of the matter
which is needed for decoupling is that the trace of the
projection of the energy-momentum tensor to the orthogonal
complement of the orbits should vanish.

\vskip 1cm\noindent
{\bf 2. CURVATURE SINGULARITIES}

The curvature of a general plane-symmetric spacetime will now
be computed. It is always possible to introduce local coordinates
so that the metric takes the form
$$ds^2=g_{ab}dx^adx^b+r^2\delta_{AB}dy^Ady^B\eqno(7)$$
Here lower and upper case indices take the values $0,1$ and
$2,3$ respectively. Let $K$ denote the Gaussian curvature of
the two-dimensional metric $g_{ab}$ and let $\nabla_a$ denote
the covariant derivative associated to that metric. Then
the curvature components are
$$\eqalignno{
R^a_{bcd}&=K(\delta^a_c g_{bd}-\delta^a_d g_{bc})&(8)     \cr
R^A_{BCD}&=-\nabla^a r\nabla_ar(\delta^A_C\delta_{BD}-
\delta^A_D\delta_{BC})&(9)                                \cr
R^a_{BcD}&=-r\nabla^a\nabla_c r\delta_{BD}&(10)}$$
Hence
$$R^{\alpha\beta\gamma\delta}R_{\alpha\beta\gamma\delta}
=4K^2+4r^{-4}(\nabla^ar\nabla_a r)^2+8r^{-2}\nabla_a\nabla_b r
\nabla^a\nabla^b r\eqno(11)$$
When the curvature components have been computed the Einstein
equations can easily be obtained. One combination of the field
equations gives
$$\nabla_a\nabla_b r=-{1\over 2r}\nabla^c r\nabla_c r g_{ab}
-4\pi r(T_{ab}-\tr T g_{ab})\eqno(12)$$
where $\tr T=g^{ab}T_{ab}$. Combining (11) and (12):
$$\eqalign{
R^{\alpha\beta\gamma\delta}R_{\alpha\beta\gamma\delta}
&=4K^2+4r^{-4}(\nabla^ar\nabla_a r)^2
+r^{-4}\left({1\over 2r}\nabla^c r\nabla_c r
-2\pi r\tr T\right)^2                         \cr
&\qquad+16\pi^2 r^{-2}\left(T_{ab}-{1\over2}\tr Tg_{ab}\right)
\left(T^{ab}-{1\over2}\tr Tg^{ab}\right)}\eqno(13)$$
The first three terms on the right hand side of (13) are
obviously non-negative and when the matter content of spacetime
is described by a massless scalar field the fourth term is
non-negative. (This condition also holds for many other
physically reasonable matter fields but that fact is not
relevant for this paper.) It follows that if
$m=-r\nabla_a r\nabla^a r/2$ then
$$R^{\alpha\beta\gamma\delta}R_{\alpha\beta\gamma\delta}
\ge 16m^2/r^6\eqno(14)$$
Returning from these general considerations to the particular
class of spacetimes considered here, it turns out that in
that case $m=\f18 e^{-\Omega}$ and $r$ is a constant times
$t^{1/2}$. It follows from (3) that $\Omega$ is non-decreasing.
Hence the curvature invariant
$R^{\alpha\beta\gamma\delta}R_{\alpha\beta\gamma\delta}$
blows up at least as fast as $t^{-3}$ as $t=0$ is approached.

It has now been shown that $t=0$ is a curvature singularity
and thus the spacetime cannot be extended further. This gives
a statement which might be called \lq strong cosmic censorship
in the past\rq\ for the class of spacetimes considered here.
(This terminology assumed that a time orientation of spacetime
has been chosen so that $t$ increases towards the future.)
It says that if a spacetime of this type is the maximal
globally hyperbolic development of initial data on some
hypersurface then no extension of that spacetime contains a
point to the past of the initial hypersurface which does not
belong to the original spacetime.

\vskip 1cm\noindent
{\bf 3. CRUSHING SINGULARITIES}

The mean curvature of the hypersurfaces of constant $t$ is
given by
$$H=-\f12 t^{1/4}e^{-\Omega/2}(\Omega_t+\f3{2t})\eqno(15)$$
Equation (3) shows that $\Omega_t\ge0$ and so (15) implies
that $|H|\ge \f34t^{-3/4}e^{-\Omega/2}\ge Ct^{-3/4}$. Thus
it can be seen that $H$ tends uniformly to $-\infty$ as the
singularity is approached. This means that this singularity is
a crushing singularity[1]. A crushing singularity in a
spatially compact spacetime always has a neighbourhood
which can be foliated by hypersurfaces of constant mean
curvature. The proof of this will now be recalled. Note
first that a spacetime which has a compact Cauchy hypersurface
can contain at most one compact spacelike hypersurface with
a given non-zero constant mean curvature[6]. The fact that
$|H|$ tends uniformly to infinity shows that given any
real number $H_1$ which is sufficiently large and negative
there exist $t_1, t_2 >0$ such that the hypersurface $t=t_1$
has mean curvature less than $H_1$ and the hypersurface
$t=t_2$ has mean curvature greater than $H_1$. These
hypersurfaces provide barriers which ensure that there
exists a hypersurface of constant mean curvature $H_1$
between the hypersurfaces of $t=t_1$ and $t=t_2$
[7,8]. Thus there is an interval $(-\infty, H_0]$ such that
the spacetime contains exactly one compact hypersurface of constant
mean curvature $H_1$ for each $H_1$ in this interval. It remains to
show that these hypersurfaces cover a neighbourhood of the
singularity. A standard result[6] implies that if $H_2<H_1$
the hypersurface of mean curvature $H_2$ lies strictly to the
past of that with mean curvature $H_1$. By construction the
hypersufaces tend to the singularity as $H_1\to-\infty$.
In other words there is no point which lies to the past of
all these hypersurfaces. It remains to show that there are
no gaps, i.e. that there is no point which lies to the
past of the hypersurface with mean curvature $H_1$ but
to the future of the hypersurfaces with mean curvature $H_2$
for all $H_2<H_1$. Suppose that a point $p$ with this property
existed. Then there would be an open neighbourhood $U$ of the
hypersurface with mean curvature $H_1$ disjoint from the
future of $p$. In $U$ there exist hypersurfaces of constant mean
curvature $H_2$ for all $H_2$ in some interval $[H_1-\epsilon,
H_1+\epsilon]$ with $\epsilon>0$. Hence there is a point
of the hypersurface with mean curvature $H_1+\epsilon$ which
lies to the past of the hypersurface with $H_1-\epsilon/2$,
contradicting a statement made earlier. It follows that no
point $p$ with the above property can exist.

\vskip 1cm\noindent
{\bf 4. VELOCITY DOMINATED SINGULARITIES}

The central problem in analysing the singularities in the class
of spacetimes considered here is to determine the behaviour
of a general spatially periodic solution of equation (2)
as $t\to 0$. Letelier and Tabensky[9] have written
down an integral formula for solutions of this equation but
they give an explicit example of a solution which is not of that
form. They conjecture that all solutions can be obtained as
limits of solutions given by the integral formula. Without a
proof of this conjecture their analysis is incomplete. This
problem can be circumvented by the direct use of energy estimates,
as has been shown by Isenberg and Moncrief[10] in the course of
a study of polarised Gowdy spacetimes. A sketch of the
argument will now be given. It will be supposed for simplicity
that the solution $\phi$ being considered is $C^\infty$. A
computation gives the inequality
$${d\over dt}\int t^2(\phi_t^2+\phi_x^2)\ge 0\eqno(16)$$
for $t>0$ when $\phi$ is a solution of (2). Since the coefficients
in the equation do not depend explicitly on the spatial coordinate
the derivative of $\phi$ of any order with respect to $x$ satisfies
the same equation as $\phi$ itself. Hence all spatial derivatives
of $\phi$ satisfy inequalities analogous to (16). The Sobolev
embedding theorem then implies that $t\phi_t$, $t\phi_x$ and the
derivatives of these quantities with respect to $x$ of any order are
bounded in a neighbourhood of $t=0$. Equation (2) can be rewritten as
$$(t\phi_t)_t=t\phi_{xx}\eqno(17)$$
Knowing that $t\phi_{xx}$ is bounded allows us to conclude that
$t\phi_t$ has a continuous extension to $t=0$. Integrating twice
in time gives the asymptotic expansion
$$\phi(t,x)=\pi(x)\log t+\omega(x)+O(t)\eqno(18)$$
for some smooth functions $\pi(x)$ and $\omega(x)$ as $t\to 0$. The
expression obtained by formally differentiating (18) once with
respect to $t$ and as many times as desired with respect to $x$ is
also a valid asymptotic expansion. Substituting these asymptotic
expansions into the evolution equation for $\Omega$ gives
$$\Omega_t=\pi^2(x)t^{-1}+O(1).\eqno(19)$$
Integrating this with respect to $t$ gives
$$\Omega(t,x)=\pi^2(x)\log t+\alpha(x)+O(t).\eqno(20)$$
for some function $\alpha$.
Let the parts of the right hand sides of (18) and (20) explicitly
written out be denoted by $\tilde\phi$ and $\tilde\Omega$
respectively so that $\phi=\tilde\phi+O(t)$ and $\Omega=
\tilde\Omega+O(t)$. The quantities $\tilde\phi$ and
$\tilde\Omega$ are solutions of the equations obtained from the
Einstein evolution equations by dropping all spatial derivatives.
This is what Isenberg and Moncrief [10] call the velocity
dominated system. Thus the solutions of the full Einstein
equations are approximated asymptotically near the singularity
by solutions of the velocity dominated system and these spacetimes
have what Isenberg and Moncrief call the AVTDS property
(asymptotically velocity-term dominated near the singularity).
This is not literally the same as the original definition of
velocity dominated singularities which was given by Eardley,
Liang and Sachs [2] but the spirit is the same and so for
brevity this property is described here as the property
that the singularity is velocity dominated.

The definition of a velocity dominated singularity makes use
of a preferred foliation by spacelike hypersurfaces. A
singularity which has the velocity dominated property
with respect to one foliation will in general not have it with
respect to a different foliation. In the present case it
has been shown that the property holds with respect to the
foliation defined by the time coordinate $t$ and this could
be interpreted as saying that this foliation is in some
sense well-behaved near the singularity. Taking this view
it is natural to ask whether the foliation by hypersurfaces
of constant mean curvature, whose existence was shown in
Sect. 3, is also well-behaved in this sense. Despite the
excellent control over the spacetime which is available,
this question appears difficult to decide. It would be
interesting to know the answer for the following reason.
The time coordinate $t$ is defined in terms of the symmetry
of the solution and so has no obvious analogue in general
spacetimes with less symmetry. On the other hand the constant
mean curvature condition makes sense in any spacetime and
it seems reasonable to hope that foliations of constant mean
curvature exist in a wide class of spacetimes.

\vskip 1cm\noindent
{\bf 5. ISOTROPIC SINGULARITIES}

In the literature there has been some discussion of isotropic
singularities, a class of singularities which is of relevance
to Penrose's Weyl curvature hypothesis. A singularity of this
kind can be defined [3, 11] by the condition that it should be
possible to conformally rescale the given metric so that the
rescaled metric extends regularly through the singularity.
In general the asymptotic form of the spacetime metric near the
singularity in the class of spacetimes considered here is
$$t[t^{\pi^2-3/2}e^\alpha(-dt^2+dx^2)+(dy^2+dz^2)]\eqno(21)$$
This shows that if $\pi^2=3/2$ everywhere the conformal class
of the metric extends continuously to $t=0$. Thus the singularity
is isotropic in this case.

In fact it is desirable to require a little more of an isotropic
singularity than what has just been demonstrated. The conformally
rescaled metric should extend not just in a continuous non-degenerate
manner to the singularity. It should also have some degree of
differentiability there. This question of the differentiability
of the rescaled metric (or more precisely the question of the
simultaneous differentiability of the rescaled metric and the
conformal factor) is in general somewhat subtle [11]. However
it turns out that in the present case everything can be made
$C^\infty$. To show this it is necessary to extend the
asymptotic expansions (17) and (19) to all powers of $t$.
An asymptotic expansion of this type for the solution of
(17) has been given in [12]. This is an expansion in
integral powers of $t$ and $\log t$. However, if $\pi$ is constant
the only term containing a logarithm is that written out explicitly
in (17). It follows immediately that in the case $\pi$=const.
(19) can be extended to an asymptotic expansion to all orders which
except for the first term is an expansion in integral powers of $t$.
This shows that the rescaled metric is $C^\infty$ in the case
identified as being isotropic above.

\vskip 1cm\noindent
{\bf REFERENCES}

\noindent
1. Eardley, D., Smarr, L. (1979) Time functions in numerical
relativity: marginally bound dust collapse. Phys. Rev. D{\bf 19},
2239-2259.
\next
2. Eardley, D., Liang, E. and Sachs, R. (1972)
Velocity-dominated singularities in irrotational dust cosmologies.
J. Math. Phys. {\bf 13}, 99-106.
\next
3. Goode, S., Wainwright, J. (1985) Isotropic singularities in
cosmological models. Class. Quantum Grav. {\bf 2}, 99-115.
\next
4. Tabensky, R., Taub, A. H. (1973) Plane symmetric self-gravitating
fluids with pressure equal to energy density. Commun. Math. Phys.
{\bf 29}, 61-77.
\next
5. Rendall, A.D. Crushing singularities in spacetimes with spherical
and plane symmetry. Unpublished.
\next
6. Marsden, J., Tipler, F. (1980) Maximal hypersurfaces and surfaces
of constant mean curvature in general relativity. Phys. Rep. {\bf 66},
109-139.
\next
7. Gerhardt, C. (1983) H-surfaces in Lorentzian manifolds. Commun.
Math. Phys. {\bf 89}, 523-553.
\next
8. Bartnik, R. (1984) Existence of maximal hypersurfaces in
asymptotically flat spacetimes. Commun. Math. Phys. {\bf 94}, 155-175.
\next
9. Letelier, P., Tabensky, R. (1975) Singularities for fluids with
$p=\omega$ equation of state. J. Math. Phys {\bf 16}, 8-10.
\next
10. Isenberg, J., Moncrief, V. (1990) Asymptotic behaviour of the
gravitational field and the nature of singularities in Gowdy
spacetimes. Ann. Phys. {\bf 199}, 84-122.
\next
11. Tod, K. P. (1992) Isotropic singularities. Rend. Sem. Mat. Univers.
Politecn. Torino {\bf 50}, 69-92.
\next
12. Chru\'sciel, P. T., Isenberg, J. and Moncrief, V. (1990) Strong
cosmic censorship in polarised Gowdy spacetimes. Class. Quantum Grav.
{\bf 7}, 1671-1680.

\end